\documentclass[conference]{IEEEtran}
\usepackage{graphicx}
\usepackage[utf8]{inputenc}
\usepackage{textcomp}

\usepackage{booktabs}
\usepackage{rotating}
\usepackage{multirow}
\usepackage{subfigure}

\usepackage{amsmath}

\usepackage{color}
\usepackage[usenames,dvipsnames,svgnames,table]{xcolor}

\ifCLASSINFOpdf
\else
\fi
\usepackage{amsmath}
\usepackage{url}

\begin{document}
%
\title{Adatom diffusion in high electric fields}

\author{\IEEEauthorblockN{Ville Jansson, Ekaterina Baibuz, Andreas Kyritsakis and Flyura Djurabekova}
\IEEEauthorblockA{Helsinki Institute of Physics and Department of Physics,\\ 
P.O. Box 43 (Pehr Kalms gata 2), FI-00014 University of Helsinki, Finland\\ Email: ville.b.c.jansson@gmail.com}}


%


\maketitle

\begin{abstract}
Strong electric fields are known to create biased adatom migration on metallic surfaces. We present a Kinetic Monte Carlo model that can simulate adatom migration on a tungsten (W) surface in electric fields. We validate our model by using it to calculate the drift velocity of the adatom at different fields and temperature and comparing the results with experimental data from the literature. We obtain excellent agreement.
\end{abstract}


%
\IEEEpeerreviewmaketitle

\section{Introduction}

It has been seen that surface protrusions or at least sharp surface feature can develop in strong electric fields \cite{nagaoka2001field,binh1992electron}. Such asperities, if large enough, may be the cause for vacuum breakdowns that occurs even in ultra high vacuum environments, as they may enhances the local field enough to cause field emission \cite{nagaoka2001field,navitski2013field,muranaka2011situ}. The bias effect of electric fields have been extensively studied both experimentally and theoretically by S. C. Wang and T. T. Tsong, who studied how W adatoms moved on flat facets of W tips in different fields with field-ion microscope \cite{tsong1975direct,wang1982field}. They were able to measure how the drift velocity of the adatoms depends on different fields. In this paper we will study the atom migration in electric fields using a Kinetic Monte Carlo (KMC) model.

\section{Methods}

For simulating the surface diffusion of atoms, we use the KMC code \textit{Kimocs}, which is described in detail in Ref. \cite{jansson2016long}. The code uses a rigid lattice, where the atom migration jumps are characterized by the number of first- and second-nearest neighbour atoms. The migration events, where an atom may jump to any unoccupied first-nearest neighbour lattice position, are chosen according to the general KMC algorithm \cite{fichthorn1991theoretical,bortz1975new,young1966monte}. The transition rates are calculated according to the Arrhenius formula
\begin{equation}\label{eq:arrhenius}
\Gamma = \nu\exp \left(\frac{-E_m}{k_B T}\right),
\end{equation}
where $\nu = 3.35\cdot10^{13}$ s$^{-1}$ is the attempt frequency, taken to be the same for all atom jump events; $k_B$ is the Boltzmann constant, $T$ is the temperature of the system, and $E_m$ is the migration energy barrier. The time increments are calculated according to the resident time algorithm \cite{bortz1975new}. In our model, the electric field above an arbitrarily rough, but still continuous, metallic surface is calculated using the field solver developed by M. Veske et al. \cite{veske2016atomistic}, which solves Laplace's equation
\begin{equation}\label{eq:laplace}
 \nabla^2 \Phi = 0,
\end{equation}
where $\Phi$ is the electrostatic potential \cite{jackson1975classical}. 

In our KMC model, the migration energy barrier $E_m$ of a jump is modified by the field according to T. T. Tsong and G. Kellogg's formula \cite{tsong1975direct}
\begin{equation}
E_m = E -(\mu_s F_s - \mu_0 F_0) - \frac{1}{2}(\alpha_s F_s^2 - \alpha_0 F_0^2),
\end{equation}
where $E$ is the migration energy barrier in the absence of any field. On the W\{110\} surface, only one barrier is possible. We use the value $E = 0.90$ eV \cite{kellogg1978direct}. $F_0$ is the field at the initial lattice position and $F_s$ is the field at the saddle point of the jump. $\mu$ is the surface-induced dipole moment and $\alpha$ is the polarizability; these values may be different at equilibrium lattice points and saddle points (as indicated by the subscripts 0 and $s$). 

\section{Results and discussion}
We verify the model for migration under electric fields by comparing the adatom drift velocity on W\{110\} surfaces under different applied electric fields and compare with the experimental data by Wang and Tsong \cite{wang1982field}. The experimental conditions were simulated by having a 3.0 nm high hemisphere with a radius of 7.44 nm made of W atoms in a body-centred-cubic lattice on a $28\times28$ nm$^{2}$ W\{110\} substrate surface. The top \{110\} facet of the hemisphere has a diameter of 2.22 nm. For an anode field of $2.35\cdot10^{10}$ Vm$^{-1}$ at the centre of the facet, a gradient of ($1.57\cdot10^{18} \pm 0.1\cdot10^{18})$ Vm$^{-2}$ between the centre and the facet edge is obtained, in good agreement with the values reported for the experiment (They report a gradient of $1.55\cdot10^{18}$ V m$^{-2}$ for the same central field \cite{wang1982field}).

\begin{figure}
\includegraphics[width=\columnwidth]{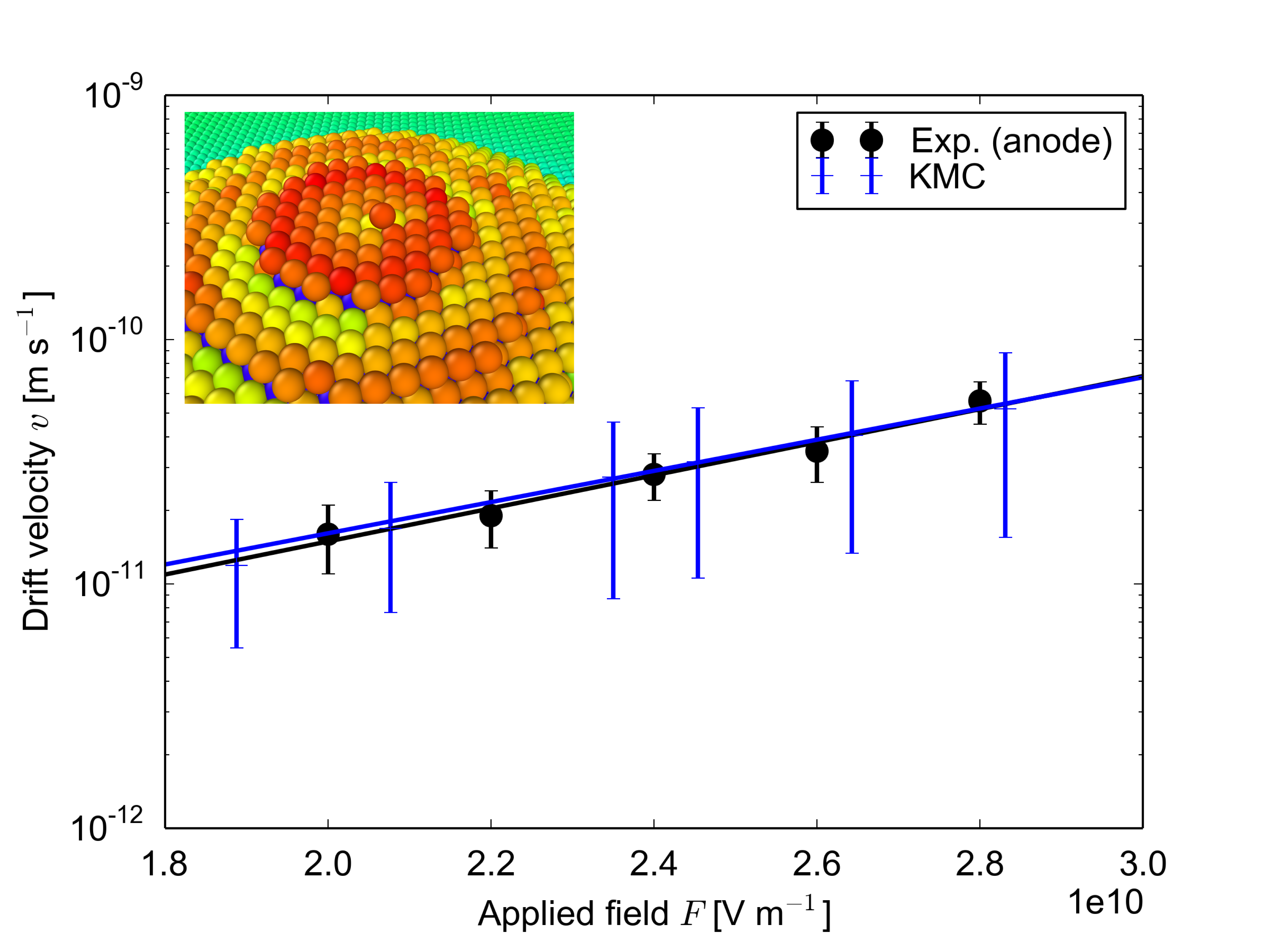}
\caption{The drift velocity for different applied fields on a W\{110\} surface; compared with the experimental data by Wang and Tsong \cite{wang1982field}. The embedded picture shows the adatom on the facet.} 
\label{graph_v_F.png}
\end{figure}

The average drift velocity was calculated by having an adatom migrating on the facet, starting one lattice position off the centre. After 3 KMC steps, the drift velocity was calculated by dividing the displacement by the time. This was repeated ten times for every applied field between $5.66\cdot10^9$ and $3.78\cdot10^{10}$ Vm$^{-1}$. The temperature was set to 280.0 K, as in the experiment \cite{wang1982field}. Using $\mu_0 = 3.72\cdot10^{-11}$ em ($5.96\cdot10^{-30}$ Cm) and $\alpha_0 = 2.39\cdot10^{-21}$ em$^2$V$^{-1}$ ($3.83\cdot10^{-40}$ Cm$^2$V$^{-1}$), which were both calculated using DFT, and $\mu_s = 3.93\cdot10^{-11}$ em, which is fitted, we obtain excellent agreement with the experimental data, as shown in Fig. \ref{graph_v_F.png}. We are here assuming that the polarizability is the same at both the lattice position and the saddle position, i.e. $\alpha_0 = \alpha_s = \alpha$. 

\begin{figure}
\includegraphics[width=\columnwidth]{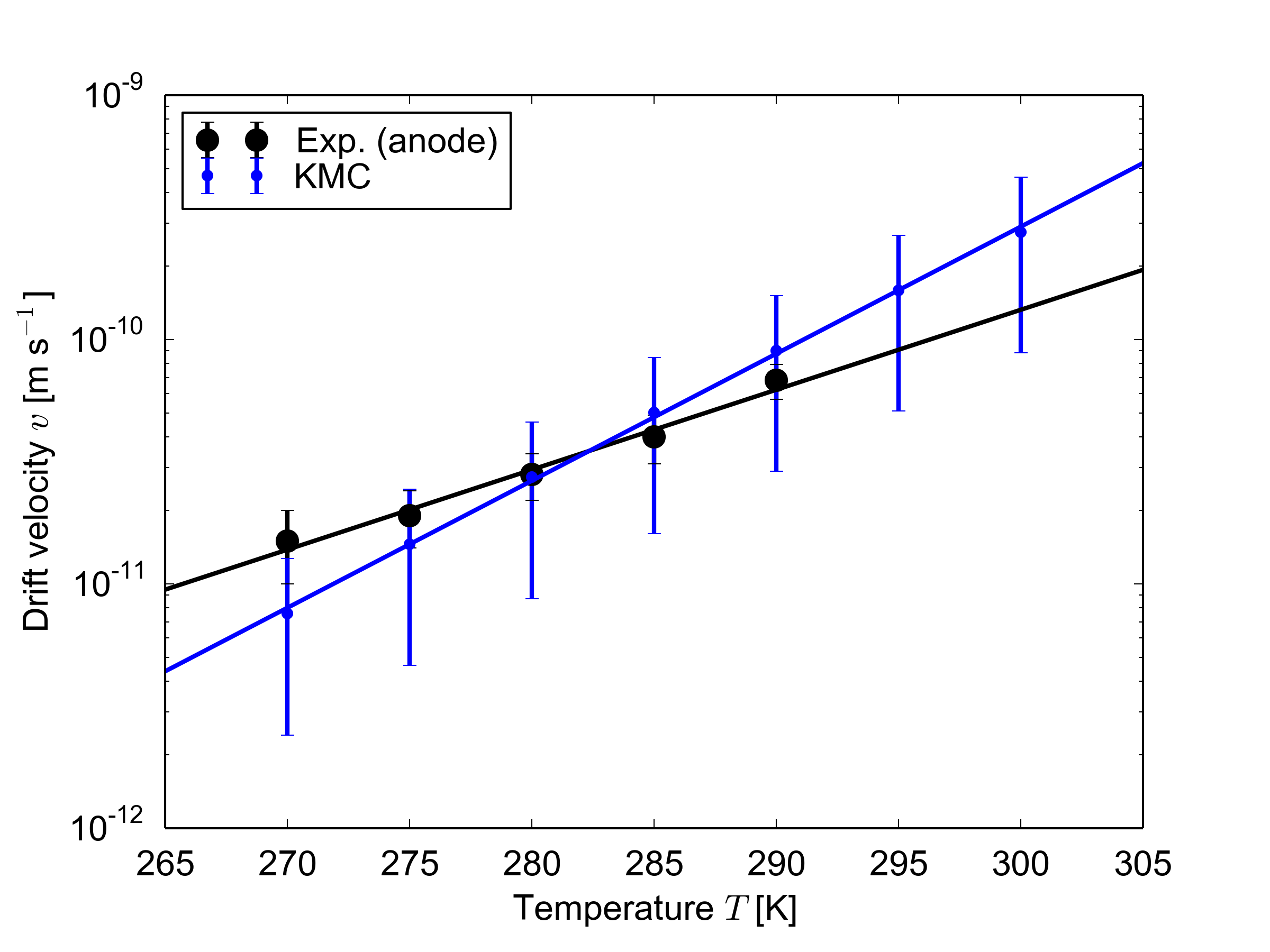}
\caption{The drift velocity for different temperatures on a W\{110\} anode surface; compared with corresponding experimental data by Wang and Tsong \cite{wang1982field}. } 
\label{graph_v_T.png}
\end{figure}

We also calculated the drift velocity at different temperatures between 270 to 290 K while keeping the applied field constant at $2.35\cdot10^{10}$ V m$^{-1}$. The results are shown in Fig. \ref{graph_v_T.png} and show good agreement with the experimental data \cite{wang1982field}.

\section{Conclusion}

We have presented a model for simulating surface diffusion of adatoms under electric fields. We have shown that our model gives excellent agreement with earlier experimental measurements of the drift velocity of W adatom diffusion on a closed-packed W\{110\} surface in different applied electric fields and at different temperatures.


\section*{Acknowledgement}

The authors acknowledges support by Academy of Finland (Grant No.\;285382 and No. 269696), CERN, and Svenska Kulturfonden (Arvid och Greta Olins Fond). Computing resources were provided by the Finnish IT Center for Science (CSC).



\bibliographystyle{IEEEtran}
%
%
%
%
%

\bibliography{/home/vjansson/Projects/vjansson,/home/vjansson/Projects/vjansson_publications}

\begin{thebibliography}{10}
\providecommand{\url}[1]{#1}
\csname url@samestyle\endcsname
\providecommand{\newblock}{\relax}
\providecommand{\bibinfo}[2]{#2}
\providecommand{\BIBentrySTDinterwordspacing}{\spaceskip=0pt\relax}
\providecommand{\BIBentryALTinterwordstretchfactor}{4}
\providecommand{\BIBentryALTinterwordspacing}{\spaceskip=\fontdimen2\font plus
\BIBentryALTinterwordstretchfactor\fontdimen3\font minus
  \fontdimen4\font\relax}
\providecommand{\BIBforeignlanguage}[2]{{%
\expandafter\ifx\csname l@#1\endcsname\relax
\typeout{** WARNING: IEEEtran.bst: No hyphenation pattern has been}%
\typeout{** loaded for the language `#1'. Using the pattern for}%
\typeout{** the default language instead.}%
\else
\language=\csname l@#1\endcsname
\fi
#2}}
\providecommand{\BIBdecl}{\relax}
\BIBdecl

\bibitem{nagaoka2001field}
K.~Nagaoka, H.~Fujii, K.~Matsuda, M.~Komaki, Y.~Murata, C.~Oshima, and
  T.~Sakurai, ``Field emission spectroscopy from field-enhanced
  diffusion-growth nano-tips,'' \emph{Applied surface science}, vol. 182,
  no.~1, pp. 12--19, 2001.

\bibitem{binh1992electron}
V.~T. Binh and N.~Garcia, ``On the electron and metallic ion emission from
  nanotips fabricated by field-surface-melting technique: experiments on w and
  au tips,'' \emph{Ultramicroscopy}, vol.~42, pp. 80--90, 1992.

\bibitem{navitski2013field}
A.~Navitski, S.~Lagotzky, D.~Reschke, X.~Singer, and G.~M{\"u}ller, ``Field
  emitter activation on cleaned crystalline niobium surfaces relevant for
  superconducting rf technology,'' \emph{Physical Review Special
  Topics-Accelerators and Beams}, vol.~16, no.~11, p. 112001, 2013.

\bibitem{muranaka2011situ}
T.~Muranaka, K.~Leifer, T.~Blom, and V.~Ziemann, ``In-situ experiments of
  vacuum discharge using scanning electron microscopes,'' Tech. Rep., 2011.

\bibitem{tsong1975direct}
T.~Tsong and G.~Kellogg, ``Direct observation of the directional walk of single
  adatoms and the adatom polarizability,'' \emph{Physical Review B}, vol.~12,
  no.~4, p. 1343, 1975.

\bibitem{wang1982field}
S.~Wang and T.~Tsong, ``Field and temperature dependence of the directional
  walk of single adsorbed w atoms on the w (110) plane,'' \emph{Physical Review
  B}, vol.~26, no.~12, p. 6470, 1982.

\bibitem{jansson2016long}
\BIBentryALTinterwordspacing
V.~Jansson, E.~Baibuz, and F.~Djurabekova, ``{Long-term stability of Cu surface
  nanotips},'' \emph{Nanotechnology}, vol.~27, no.~26, p. 265708, 2016.
  [Online]. Available: \url{http://arxiv.org/abs/1508.06870}
\BIBentrySTDinterwordspacing

\bibitem{fichthorn1991theoretical}
K.~A. Fichthorn and W.~H. Weinberg, ``Theoretical foundations of dynamical
  monte carlo simulations,'' \emph{The Journal of Chemical Physics}, vol.~95,
  p. 1090, 1991.

\bibitem{bortz1975new}
A.~B. Bortz, M.~H. Kalos, and J.~L. Lebowitz, ``{A new algorithm for Monte
  Carlo simulation of Ising spin systems},'' \emph{Journal of Computational
  Physics}, vol.~17, no.~1, pp. 10--18, 1975.

\bibitem{young1966monte}
W.~M. Young and E.~W. Elcock, ``{Monte Carlo studies of vacancy migration in
  binary ordered alloys: I},'' \emph{Proceedings of the Physical Society},
  vol.~89, p. 735, 1966.

\bibitem{veske2016atomistic}
M.~Veske, A.~Kyritsakis, F.~Djurabekova, R.~Aare, K.~Eimre, and V.~Zadin,
  ``Atomistic modeling of metal surfaces under high electric fields: Direct
  coupling of electric fields to the atomistic simulations,'' in \emph{Vacuum
  Nanoelectronics Conference (IVNC), 2016 29th International}.\hskip 1em plus
  0.5em minus 0.4em\relax IEEE, 2016, pp. 1--2.

\bibitem{jackson1975classical}
J.~Jackson, \emph{Classical electrodynamics}.\hskip 1em plus 0.5em minus
  0.4em\relax Wiley, New York, 1975.

\bibitem{kellogg1978direct}
G.~Kellogg, T.~Tsong, and P.~Cowan, ``Direct observation of surface diffusion
  and atomic interactions on metal surfaces,'' \emph{Surface Science}, vol.~70,
  no.~1, pp. 485--519, 1978.

\end{thebibliography}

\end{document}